\documentclass[useAMS,usenatbib]{mn2e}

\usepackage{psfig}

\title[Rees-Sciama effect]{On the Rees-Sciama effect: maps and statistics}
\author[ N. Puchades et al.]{N. Puchades$^{1}$ M.J. Fullana$^{2}$ 
J.V. Arnau$^{3}$ and D. S\'aez$^{1}$\footnotemark[0]\thanks{
E-mail: diego.saez@uv.es (DS)}\\
$^{1}$Departamento de Astronom\'{\i}a y 
Astrof\'{\i}sica, Universidad de Valencia.
46100 Burjassot, Valencia, Spain\\
$^{2}$Institut de Matem\`atica Multidisciplinar,
Universitat Polit\`ecnica de Val\`encia. 46022 Val\`encia,
Spain\\
$^{3}$Departamento de Matem\'atica Aplicada,
Universidad de Valencia.
46100 Burjassot, Valencia, Spain 
}

\begin{document}

%\date{Accepted 2006 xxxxx yy. Received 2006 April 20; in 
%original form 2006 january 30}

\pagerange{\pageref{firstpage}--\pageref{lastpage}} \pubyear{2005}

\maketitle

\label{firstpage}

\begin{abstract}

Small maps of the Rees-Sciama (RS) effect are simulated
by using an appropriate N-body code and a certain ray-tracing procedure.
A method designed
for the statistical analysis of cosmic microwave background (CMB) maps
is applied to study the resulting simulations.
These techniques, recently proposed --by our team-- to consider lens 
deformations of the CMB, are adapted to deal with the RS effect.
This effect and the deviations from Gaussianity associated to it seem
to be too small to be 
detected in the near future. This conclusion follows from our
estimation of both the RS angular power spectrum and the RS 
reduced $n$-direction
correlation functions for $n \leq 6$.  

\end{abstract}


\section{Introduction}

From decoupling to present time,  
Cosmic Microwave Background (CMB) photons have been traveling
--almost freely-- through a homogeneous and isotropic background 
universe with small perturbations.
Taking into account the main results from the analysis
of the first year WMAP (Wilkinson Microwave Anisotropy Probe) data,
which were presented in \citet{ben03},
it is assumed: 
(i) a Gaussian distribution 
of adiabatic energy density fluctuations with
Zel'dovich spectrum, (ii) 
a currently accelerated expansion with reduced Hubble constant 
$h=10^{-2}H_{0}=0.71$ ($H_{0}$ being the 
Hubble constant in units of $Km \ s^{-1} Mpc^{-1}$), (iii)
a cosmological constant
with density parameter $\Omega_{\Lambda} = 0.73$ producing 
the current acceleration,
and (iv) cold dark matter (baryonic matter) with  
density parameter $\Omega_{d} =0.23$ ($\Omega_{b} = 0.04$);
hence, the total matter density parameter is $\Omega_{m} = \Omega_{b}
+ \Omega_{d} = 0.27$.
In other words, a realization of the so-called concordance model  
is assumed and, then, 
the CMBFAST code \citep{selz96} is used to get
the linearly evolved power spectrum, 
$P(k)$, of the 
scalar energy density fluctuations, which is defined 
by the relation $\langle \delta (\vec{k}) \ \delta^{*} (\vec{k}^{\prime}) 
\rangle = (2 \pi )^{3} P(k) \delta (\vec{k} - \vec{k}^{\prime}) $. 
This spectrum is normalized 
with the condition $\sigma_{8} = 0.93$ to have 
the required density of galaxy clusters \citep{eke96}. 
 
Along this paper, 
units are chosen in such a way that the speed of light is $c=1$ and
the gravitation constant is $G=1/8\pi $.
Whatever quantity "$A$" may be, $A_{_{L}}$, $A_{0}$, and  $A_{_{B}}$ 
stand for
the $A$ values on 
the last scattering surface, at present time, and in the background,
respectively. Symbols 
$\vec {x}$, $a$, $z$, and $\eta$,
stand for the comoving position vector, the scale factor, the redshift,
and the conformal time, respectively.
Condition $a_{0}=1 \ Mpc$ is assumed.

\begin{figure}
\psfig{file=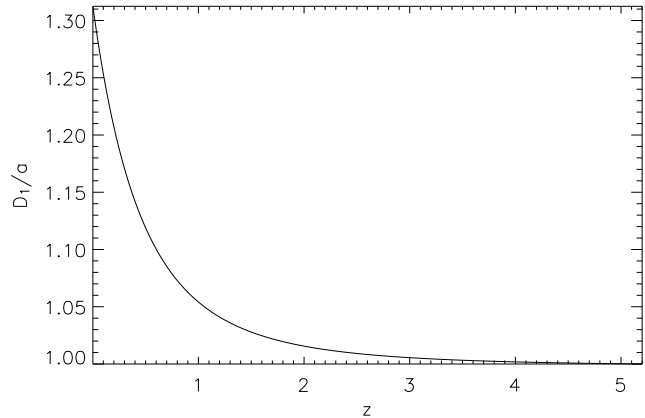,width=8.4cm}
%\vspace{0.5cm}
%% to leave a blank space
\caption{Ratio $D_{1} /a$ as a function of $z$ in the concordance
model}
\label{figu0}              
\end{figure}

Cosmological structures (scalar modes) produce a 
peculiar gravitational potential $\phi (\vec{x},\eta )$
which interacts with the CMB.
The partial time derivative of
$\phi (\vec{x}, \eta )$
produces temperature fluctuations.
Quantity $\Delta T/T_{_{B}} = (T-T_{_{B}})/T_{_{B}}$  
measures fluctuations with respect to the 
CMB averaged temperature $T_{_{B}}$. This
quantity is given by the formula:
\begin{equation}
\frac {\Delta T}{ T_{_{B}} } (\vec {n})= 2 \int_{\eta_{_{L}}}^{\eta_{0}} 
\frac {\partial \phi (\vec{x},\eta )} {\partial \eta } d \eta   \ ,
\label{rtp}
\end{equation}
where the integral is to be performed along the 
null geodesic of the perturbed universe associated to the observation
direction $\vec{n}$. As it is well known, 
a good approximation to $\Delta T/ T_{_{B}}$ can be obtained by 
calculating this
integral along the
background null geodesic corresponding to $\vec {n}$, which obeys the equation
\begin{equation}
x^{i}= \lambda (z) n^{i}  \ ,
\end{equation}
where
\begin{equation}
\lambda (z)= H_{0}^{-1} \int_{0}^{z} \frac {db} {[\Omega_{m}(1+b)^{3}+
\Omega_{_{\Lambda}}]^{1/2}} \ .
\end{equation}
In the flat universe with cosmological constant we are considering,
linear, mildly non-linear, and strongly non-linear structures 
contribute to the
integral in  Eq. (\ref{rtp}). Indeed, this integral allows us to
estimate the anisotropy 
produced by the total time varying peculiar gravitational potential
of the universe. This anisotropy  could be considered as an unique effect;
nevertheless,
the contributions --to the integral-- due to linear and non-linear 
structure evolution are usually called 
the Integrated Sachs-Wolfe (ISW) and the Rees-Sciama (RS) effect, 
respectively. 
Some comments about this distinction seem to be worthwhile.
During linear evolution, the peculiar potential can be written 
in the form 
$\phi_{_{L}} D_{1}(a) / a$, where $\phi_{_{L}}$ is the initial
potential (taken at decoupling time) and
$D_{1}(a)$ is the linear growth function        
for the density contrast. At any time, the total peculiar potential,
$\phi $, 
can be splitted as follows:
\begin{equation}
\phi = \phi_{_{L}} \frac {D_{1}(a)}{a} + \phi_{_{NL}} \ ,
\label{sepiden}
\end{equation}
where $\phi_{_{NL}} $ is the part of the potential properly due 
to nonlinear evolution.  
After substitution in Eq. (\ref{rtp}) and integration until present time, 
the first term of the r.h.s. of this equation gives the 
ISW effect, whereas the second one is the source of the 
RS effect. Equation (\ref{sepiden}) leads to a consistent definition of 
both effects (the addition is the total integrated 
gravitational effect). Before the beginning of the nonlinear regime, 
function $\phi_{_{NL}} $ takes on negligible second order values
and only the ISW anisotropy can be significant. During the nonlinear regime, 
the first term continues contributing to the ISW effect and the 
second term produces the RS anisotropy.    
Let us now study the ratio $D_{1}(a)/a$ to 
discuss in more detail ISW anisotropy production
in the concordance model, in which, function  
$D_{1}$ can be written as follows (in terms of $z$):
\begin{equation}
D_{1}(z) \propto \frac {1}{x} \left[ \frac {2}{x} + x^{2} \right]^{1/2}
\int_{0}^{x} \left[ \frac {2}{y} + y^{2} \right]^{-3/2} dy  \ , 
\label{ll2}       
\end{equation}
where
\begin{equation}
x = \left[ \frac {2 \Omega_{\lambda}}{\Omega_{m}} \right]^{1/3}
(1+z)^{-1} \ . 
\label{ll3}
\end{equation} 
With the appropriate normalization factor, function $D_{1}$ 
is almost identical to the scale factor $a$ for $z>3$, 
whereas it is 
greater than this factor for $z<3$. 
This behaviour is shown in Fig. 
(\ref{figu0}), where we see that function $D_{1} / a$ is very 
close to (greater than) unity for $z>3$ ($z<3$).
Two periods can be then distinguished: in the first one 
($z>3$), the term $\phi_{_{L}} D_{1}(a) / a$ is almost constant
and no significant ISW anisotropy is generated. In the
second period ($z<3$), the potential $\phi_{_{L}} D_{1}(a) / a$
varies and the ISW effect is produced.
These facts can be easily understood.
For $z>3$, vacuum energy is subdominant and the concordance 
model approximately evolves as a flat universe without 
cosmological constant; in which, it is well known that
no ISW is generated because the relation 
$D_{1}(a) \simeq a$ is satisfied.
All the ISW effect is produced when vacuum energy plays a 
significant
role in the evolution of the universe; namely, for $z<3 $;
therefore, in the concordance model, the ISW and the RS effects 
are both produced at low redshifts. 
Only a small contribution to the RS effect is produced 
at $z>3 $ as a result of the first stages
of non-linear evolution. The main part of the RS effect
and all the ISW anisotropy are simulataneously generated at redshifts
$z<3 $. The ISW (RS) effect dominates for small (large) 
enough multipoles (see next sections).

Since the location of the first peak of the CMB angular 
power spectrum indicates that the background universe 
is almost flat and, in such a case, the ISW effect only 
appears in the presence 
of vacuum energy (cosmological constant),
any detection of the ISW effect could be considered as an 
indirect detection of vacuum energy.
Since the largest ISW multipoles 
correspond to small $\ell $ values with large cosmic variances,
a direct detection based on the analysis of CMB maps 
becomes problematic
\citet{hu02}; nevertheless, ten years ago, \citet{cri96} proposed 
an indirect detection method which is 
based on the estimate of cross-correlations between the CMB 
maps and tracers of the matter density distribution 
at $z < 3 $   
(risponsable for the ISW anisotropy). Many papers
have been devoted to try this indirect detection 
(\citet{fos03, fos04, bou04, afs04, nol04, vie06});
nevertheless, in the
present paper, 
the ISW effect is only considered with the essential aim of estimating
its contribution to our maps 
of the integrated gravitational effect, which are 
obtained from Eq. (\ref{rtp}) and appropriate numerical techniques
(see below). 
Since the ISW contribution
is proved to be small enough (see next sections), 
our maps can be considered as approximated simulations of
the RS effect; by this reason, these maps are hereafter
called RS maps and the spectra extracted from them are considered 
as an approximation to the RS spectrum.

This paper is organized as follows: Appropriate comments on the
ISW effect are given in section 2. The numerical methods used to 
simulate maps of the RS effect are described in section 3. 
The RS angular power spectrum obtained from small maps is 
described in section 4, where the importance of the 
smallness of the maps
(as a source of uncertainty) is discussed in detail.
Next section deals with the estimation of both reduced
m-correlations and 
deviations from Gaussianity and, finally, section 6 contains a summary of
conclusions and a general discussion.  

\section{On the Angular Power Spectrum of the ISW effect}

\begin{figure}
\psfig{file=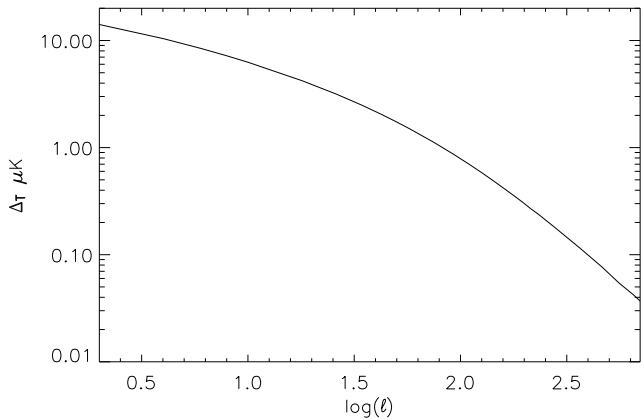,width=8.4cm}
%\vspace{0.5cm}
%% to leave a blank space
\caption{ISW power spectrum, $\Delta_{_{T}} $, in $\mu K$, as a function 
of $\log (\ell )$.}
\label{figu1}              
\end{figure}

In a flat background with cosmological constant,
the angular power spectrum of the ISW effect depends
on various parameters.
The values of $h$, $\Omega_{\Lambda}$, and 
$\sigma_{8}$ are particularly important.
This spectrum has been previously calculated in
various cases, see, e.g. \citet{hu02, coor02}.
Here, it is calculated for the cosmological model and 
parameters 
fixed in the introduction.
The $C_{\ell} $ quantities of the ISW effect can be written 
as follows \citep{ful00}:
\begin{equation}
C_{\ell}= N
\int_0^\infty \frac {P(k)} {k^{2}} \xi^{2}_{\ell} (k) dk  \ , 
\label{isw1}
\end{equation}
where
\begin{equation}
\xi_{\ell} (k) = \int_{\lambda_{0}}^{\lambda_{_{L}}} 
j_{\ell } [\lambda k]  
\frac {d} {d\lambda} \left[ (1+z)D_{1}(z) \right] d\lambda  \ ,
\label{isw2}
\end{equation}
$j_{\ell}$ being the spherical Bessel function of order $\ell$,
and
\begin{equation}
N = \frac {18H_{0}^{4}} {\pi}                                  
\left[ \frac {\Omega_{m}} {D_{1}(0)} \right]^{2} \ ,
\end{equation}
where function $D_{1}(z)$ is given by Eqs. (\ref{ll2}) and 
(\ref{ll3}).
The integrals in Eqs. (\ref{isw1}) and (\ref{isw2}) have been numerically
calculated to get the required spectrum.
Results are presented in Fig. (\ref{figu1}), 
where quantity $\Delta_{_{T}} = [ \ell (\ell +1) C_{\ell} /
2 \pi ]^{1/2} $ in $\mu K$ (used along the paper to describe
the angular power spectrum) is displayed in the $\ell $ interval
[2,700].

Our estimation, as well as previous ones
(see e.g. \citet{hu02, coo02, pad05}),
leads to two main conclusions: 
(1) the most important part of the ISW effect is produced by very 
large spatial scales (small $k$ values) contributing to small multipoles, 
and (2) the contribution to this effect strongly decreases as the 
spatial scale does. For the parameters we have assumed,
the $\Delta_{_{T}}$ value corresponding to 
$\ell = 700$ is $ 0.037 \ \mu K$
and, moreover, quantity $\Delta_{_{T}}$ rapidly decreases as $\ell $ 
increases reaching a value close to $0.0045 \ \mu K$ for
$\ell \sim 2000$. 
These values are to be 
compared with those extracted from our numerically 
simulated maps. As a technical requirement associated to ray-tracing 
(see below), our
maps are produced by the total time varying 
peculiar potential due to all the scales 
smaller than $60 \ Mpc $, which are evolved by N-body simulations
from redshift $5.2$ to present time; hence, 
all the ISW effect produced (at $z < 3$, see Fig. (\ref{figu0})) by 
these scales is included in the resulting maps.
The effect produced by scales greater than $60 \ Mpc$
must be independently calculated. Is it a pure ISW
effect? A few considerations about linearity are 
useful to answer this question.

\begin{figure}
\psfig{file=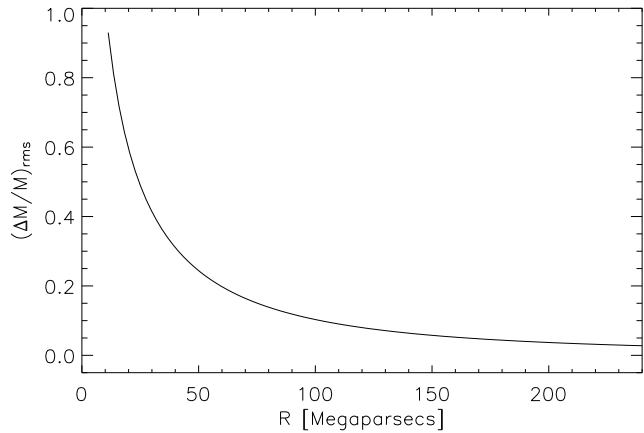,width=8.4cm}
%\vspace{0.5cm}
%% to leave a blank space
\caption{rms relative mass fluctuations inside a sphere whose present 
radius in Megaparsecs is R.
}
\label{figu2}
\end{figure}

In order to
discuss about linearity in detail, the root mean square 
(rms) of the relative mass fluctuations inside a sphere having 
a present radius of $R \ Mpc$ (randomly placed in our flat universe) 
is estimated. Condition $a_{0}=1$ implies that number $R$
is also the comoving radius of the sphere.
This rms value is $\sigma(R) = (\Delta M/M)_{rms}
= \langle (\delta M/M)^{2} \rangle^{1/2}$,
with
\begin{equation}
\langle (\delta M / M)^{2} \rangle =
\frac {1}{2\pi^{2}}\int_0^\infty k^{2} P(k) W^{2}(kR) dk  \ ,
\end{equation} 
where $W(kR)$ is the window function of the R-sphere 
[$W(kR) = \frac {3}{y^{3}} (\sin y-y \cos y)$ with
$y=kR$; see \citet{pee80}].
Numerical calculations based on these formulae
lead to the
$\sigma (R) $ values presented in Fig. \ref{figu2}. From
this figure it follows that, for the chosen 
normalization of $P(k)$ and the comoving radius $R = 15 \ Mpc$, 
the $\sigma_{15}$ 
value is close to $0.7$; hence, it can be assumed  
(standard point of view) that
regions having a comoving diameter greater than
$30 \ Mpc$ can be treated as linearly evolving zones.
The scale $\hat{L} = 60 \ Mpc$, used along this paper, 
evolves in the linear regime with $\sigma(30) \simeq 0.4$; 
hence, scales greater than $60 \ Mpc$ would produce 
a pure ISW effect to be separately estimated with 
appropriate linear techniques.
The ISW effect produced by all the scales smaller than $60 \ Mpc$
is included in our maps and, obviously, 
it is smaller than the total ISW effect 
(see Fig. \ref{figu1} and previous comments).

\section{Numerical Methods}

RS is a pure gravitational effect and, consequently,
only the dominant dark matter component is considered,
whereas the sub-dominant baryonic component is
neglected. On account of these facts,
the formation and evolution of non-linear cosmological structures
can be described by using N-body simulations with appropriate
boxes and resolutions. In the PM simulations we use here
\citep{hoc88,qui98}, the 
peculiar potential satisfies the equation \citep{mar90}:
\begin{equation}
\Delta \phi = \frac {1}{2} a^{2} (\rho_{m} - \rho_{mB}) \ .
\label{cden}
\end{equation} 
The universe is assumed to be covered by simulation boxes and, 
consequently, it is periodic. As it was pointed out by
\citet{cer04} and \citet{ant05}, periodicity effects magnify lens deformations
and, in general, gravitational anisotropies as the RS effect.
Various techniques (ray-tracing methods) have been used to avoid 
this magnification, for example, \citet{tul96} averaged the temperature
contrasts of many rays, but then, they found a map with too large pixels,
from which, the multipoles corresponding to $\ell > 700$ could not  
be obtained. Another method based on multiple plane
projections  was proposed and applied to study 
week lensing by cosmological structures
(see \citet{jai00} for a detailed description). Afterwards, 
\citet{whi01} designed the "tiling" ray tracing procedure, 
in which, independent simulations with appropriate 
boxes and resolutions tile the photon trajectories.
Finally, another method based on the existence 
of preferred directions and on the use of an appropriate 
cutoff was recently proposed 
and applied (\citet{cer04}; \citet{ant05}). In  
the required cutoff, the Fourier modes corresponding to
spatial scales larger than a given one ($L_{max}$) are eliminated 
from the peculiar gravitational potential. The $L_{max}$ value 
is chosen in such a way that: (a) the cutoff eliminates all 
the spatial scales which are too large to be well described in the
simulation box, and (b) after the cutoff is performed, CMB photons cross
neighbouring boxes through statistically independent regions. 
This cutoff is only performed
in the output peculiar gravitational potential
with the essential aim of calculating the integral 
in Eq. (\ref{rtp}); however,
all the spatial scales allowed 
by the box size are taken into account in the N-body simulation. 
Our ray-tracing method is a very good choice presenting
some advantages with respect to other techniques 
(see \citet{ant05}), e.g., there are no discontinuities at the 
crossing points
between successive boxes, the 
computational cost is moderated, and so on.

In order to study the RS effect, we develop the same type of
N-body simulations which were
used, in \citet{ant05}, to deal with lensed maps of the CMB. In 
this way, various maps of the relative 
temperature variations due to the RS
effect are created and analyzed, and the dependence of the results 
on various parameters 
involved in the simulations and also in their 
statistical analysis is then 
discussed. Most maps are regularly pixelised, but some maps based
on HEALPIx ({\em Hierarchical Equal Area Isolatitude 
Pixelisation of the Sphere}, see \citet{gor99}) 
are also considered.
Taking into account that the box size used in the mentioned 
N-body simulations 
(see \citet{ant05} for details) is $L= 256 \ Mpc$, and also that the main
part of the RS effect is assumed to be produced at redshift
$z < 5.2$ (see below), the size of the resulting 
RS maps appears to be close to $2^{\circ}$. 
Simulation boxes cover a periodic universe and, for the 
direction $\theta=77.2^\circ $ and $\phi = 12.6^\circ$ 
(where $\theta$ and $\phi$
are spherical coordinates defined with respect to the 
box edges), the CMB photons cross around $30$ boxes before
passing through the region where they were initially located
(in the first box). 
It can be verified that these boxes cover
the trajectory of the CMB photons from the redshift 
$z_{in} \simeq 5.2$ to present time, and also that 
these photons cross successive boxes through independent regions 
separated by a distance close to $52 \ Mpc$. The cutoff
is performed at the scale $k_{min} =2 \pi / L_{max}$
with $L_{max} = 60 \ Mpc$ (in agreement with previous comments
about the scales involved in our simulations).

\begin{figure}
\psfig{file=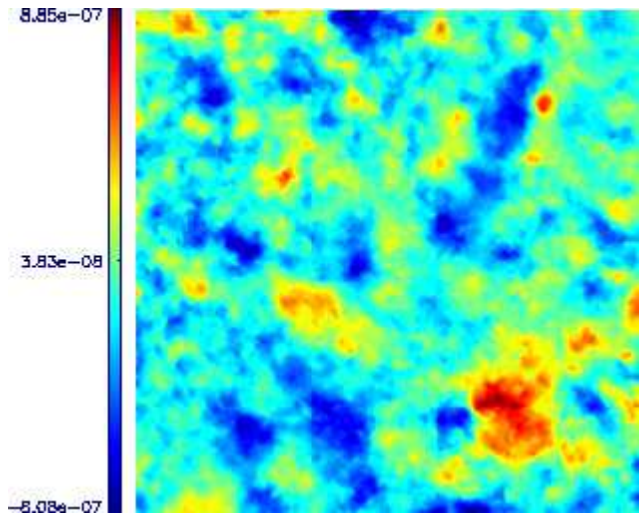,width=8.4cm}
%\vspace{0.5cm}
%% to leave a blank space
\caption{Simulated $ 1.83^{\circ} \times 1.83^{\circ} $
map of the RS effect obtained from a HR N-body
simulation}
\label{figucolour}              
\end{figure}

Along the paper we develop three types of N-body simulations
in boxes of $L \ Mpc$ (integer and even L): 
Low Resolution (LR) simulations 
involving $L/2$ cells per edge and $(L/2)^{3}$ particles
($2 \ Mpc$ cell size and $\sim 380$ time steps),
Intermediate Resolution (IR) simulations 
using L cells per edge and $L^{3}$ particles
($1 \ Mpc$ cell size and $\sim 410$ time steps), and
High Resolution (HR) simulations                   
with $2L$ cells per edge and $(2L)^{3}$ particles
($0.5 \ Mpc$ cell size and $\sim 500$ time steps).
Figure (\ref{figucolour}) shows a RS map which has been 
obtained with our ray-tracing method and a HR 
simulation of structure. Its look is promising, 
but the correlations
appearing in this type of maps must be estimated
to get objective conclusions.
It is done in next sections.

For any observation direction $\vec {n}$, function 
$\partial \phi / \partial \eta $ must be calculated in 
a set of points 
to estimate the integral in Eq. (\ref{rtp}). We have chosen
equidistant points (on the corresponding background geodesic)
separated by a comoving distance equal to
the cell size. Each of these points has a comoving 
position vector $\vec{x}_{_{P}}$, and the photon passes by
this position at time $\eta_{_{P}} $. 
Point $P$ is placed inside a certain  $P$-cell of the 
computational grid and time  $\eta_{_{P}} $ belongs to  
some interval [$\eta_{i}, \eta_{i+1} $], where 
$\eta_{i} $ and $\eta_{i+1} $ are two successive times
of the N-body simulation, at which, the peculiar
potential $\phi $ is found at any grill node. 
The potential at point $P$ is calculated at times 
$\eta_{i-1} $, $\eta_{i} $, $\eta_{i+1} $, and $\eta_{i+2} $.
The calculation at any of these times is performed by 
means of a suitable 3D interpolation, which uses the  
potential in the vertices of the 
$P$-cell and in those of the neighbouring ones.
From the resulting potential 
at the above four times, 
the partial derivative $\partial \phi / \partial \eta $
is finally calculated with an appropriated numerical method. 
For each type of N-body simulations (LR, IR or HR),
various methods
for spatial interpolation and also for the numerical calculation 
of time derivatives
have been tried. 
The final results do not significantly depend on the chosen
methods because these results depend much more on the 
simulation resolution (see comments on Fig. (\ref{figu4}) given below). 
Two types of spatial interpolations have been considered: 
(i) an standard 3D linear interpolation using the potential
in the eight nodes of the $P$ cell, and (ii) a 3D interpolation 
based on the use of splines along each direction; in this case,
four 
vertices located in the $P$-cell and in their neighbouring ones 
are considered
for interpolations in the directions of the box edges; results
from both interpolations are very similar; relative differences 
between the $C_{\ell} $ coefficients extracted from the 
maps obtained with the two interpolations 
are smaller than $0.02 $ for any $\ell $.
Also time derivatives based on different numbers of points, $\eta_{i}$,
have been performed, but the results 
are almost identical in all cases because 
$\phi $ is a very smooth function of $\eta $
in any node of the simulation grid 
(see \citet{ali02}). In short, a 3D linear interpolation in 
space and a standard derivative based on two times
$\eta_{i} $ and $\eta_{i+1} $ suffice to get good maps 
of the RS effect. Various methods have been also used 
to perform the integral in Eq. (\ref{rtp}): trapezoidal, Simpson,
and so on. Results are almost identical in all cases.
All these comments show that our ray-tracing procedure is a robust numerical 
method, which leads to RS maps having almost the same properties for
a wide 
set of numerical techniques (interpolations, derivatives and integrations).

\section{On the Angular Power Spectrum of the RS effect}
 
Since we are concerned with small squared regularly pixelised maps, 
the power spectrum estimator described in 
\citet{arn02} applies. Although this estimator is used as the basic one 
along the paper, some results are compared with those of the code ANAFAST
included in the HEALPIx package.

\begin{figure}
\psfig{file=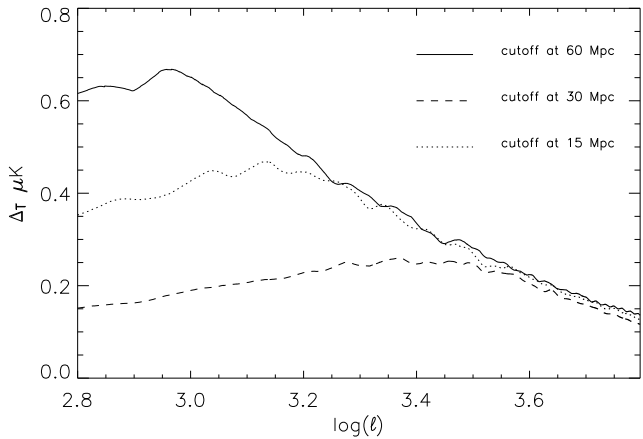,width=8.4cm}
%\vspace{0.5cm}
%% to leave a blank space
\caption{RS power spectrum, $\Delta_{_{T}} $, in $\mu K$, as a function 
of $\log (\ell )$. Solid, dotted and dashed lines are the spectra 
produced by spatial scales smaller than $60 \ Mpc$, $30 \ Mpc$,
and $15 \ Mpc$, respectively. IR simulations have been used.
}
\label{figu3}
\end{figure}

\begin{figure}
\psfig{file=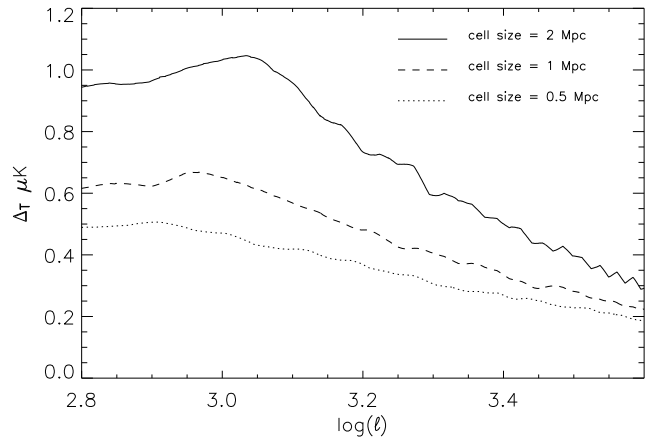,width=8.4cm}
%\vspace{0.5cm}
%% to leave a blank space
\caption{RS power spectrum, $\Delta_{_{T}} $, in $\mu K$, as a function 
of $\log (\ell )$. Solid, dashed and dotted lines are the spectra 
obtained from the LR, IR, and HR simulations defined in the text, 
respectively. Scales are smaller than $60 \ Mpc$ in all cases.
}
\label{figu4}
\end{figure}

The basic estimator is first used to analyze 
$ 1.83^{\circ} \times 1.83^{\circ} $
RS maps obtained from IR simulations with $L=256 \ Mpc$. 
Figure \ref{figu3} shows the resulting
angular power spectra ($\Delta_{_{T}}$) for three appropriate choices of
the cutoff: $60$, $30$, and $15 \ Mpc$. 
Differences between the $\Delta_{_{T}}$ values corresponding to the solid
(dotted) and dotted (dashed) lines are produced by scales between 
$60$ and $30 \ Mpc$ ($30$ and $15 \ Mpc$). The relation 
$\Delta_{_{T}} > 0.037 \ \mu K$ is satisfied and, according
to previous comments (see section 2), the ISW effect cannot
produce these spectra; therefore,
scales between $60 \ Mpc$ and $15 \ Mpc$
seem to produce a certain Rees-Sciama effect, which is only
possible if they become 
non-linear enough in our simulation boxes. The question is: are 
the scales between $15$ and $60 \ Mpc$ well described in our simulations?
In order to answer, it should be taken into account that
the evolution of the scales under consideration depend on  
N-body simulation 
characteristics as resolution and box size; indeed, 
the greater the N-body resolution, the better (more realistic) 
the description of all the spatial scales and, in particular,
the description of the scales between $15$ and $60 \ Mpc$.
We can say then that part of the non-linear behaviour of these 
scales is due to low resolution
(see next paragraph for more discussion).

For a cutoff at $60 \ Mpc$ and a fixed box size $L=256 \ Mpc$, the 
angular power spectrum of the RS maps has been calculated for HR, IR, 
and LR simulations; namely, for different space and mass resolutions.
The resulting spectra are displayed in Fig. \ref{figu4}, where we
see that the greater the resolution the smaller the amplitude
of the RS effect. Taking into account 
Fig. \ref{figu3} this behaviour of the amplitude 
is not surprising because, as resolution increases, 
the evolution of the scales between $60$ and $15 \ Mpc$ is 
better described and  
masses concentrate more and more on smaller and smaller scales.
Taking into account that the separation between the solid 
and dashed lines is greater than that of the 
dashed and dotted lines, it seems that, although 
the spectra of Fig. \ref{figu4} depend on resolution, they 
tend to a certain spectrum located below 
the dotted line and close to it; hence, this line gives an upper limit to the 
true spectrum. The difference between the true spectrum and the
upper limit (dotted line) is expected to be small and, consequently,
the dotted line can be considered as our best estimation of  
the RS angular power spectrum. In spite of this fact, there are
systematic errors in this spectrum associated to the smallness of
our simulated maps. These errors are analyzed below

\begin{figure}
\psfig{file=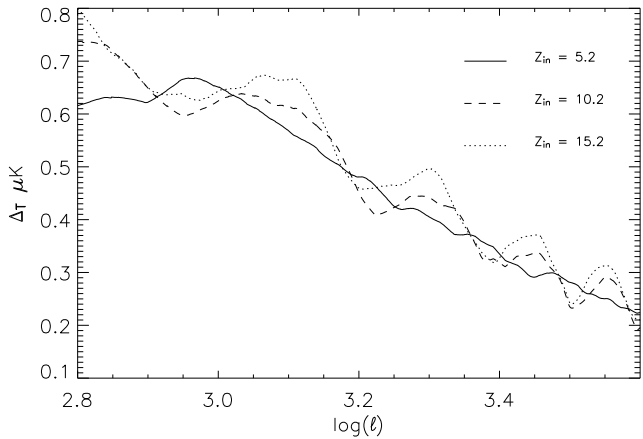,width=8.4cm}
%\vspace{0.5cm}
%% to leave a blank space
\caption{RS power spectrum, $\Delta_{_{T}} $, in $\mu K$, as a function 
of $\log (\ell )$. Solid, dashed, and dotted lines correspond to 
the spectra obtained from the initial redshifts $5.2$, $10.2$, and 
$15.2$, respectively. IR simulations and scales smaller than 
$60 \ Mpc$ have been considered in the three cases.
}
\label{figu5}
\end{figure}

The spectra of Fig. \ref{figu4} have been obtained for an initial redshift
$z_{in} = 5.2$, but other initial redshifts have been tried.
Fig. \ref{figu5} shows the spectra obtained from initial redshifts of
5.2, 10.2 and 15.2. In the three cases, IR simulations have been used
and the size of the simulated maps is $ 1.83^{\circ} \times 1.83^{\circ} $.
In order to maintain the map area for the three initial redshifts, 
the size of the simulation box (and accordingly the resolution) has 
been varied. The assumed box size (N-body resolution)
slightly increases (decreases) as the initial redshift grows. 
Figure \ref{figu5} shows 
oscillations in $\Delta_{_{T}}$ whose amplitude increases as 
$z_{in} $ does. 
In spite of these unexplained oscillations, the resulting spectra 
are rather similar in the three cases and, as it follows from 
this figure, the choice of the initial 
redshift $z_{in} = 5.2 $ is the simplest one leading to  a good estimate of 
$\Delta_{_{T}}$ without spurious oscillations.

\begin{figure}
\psfig{file=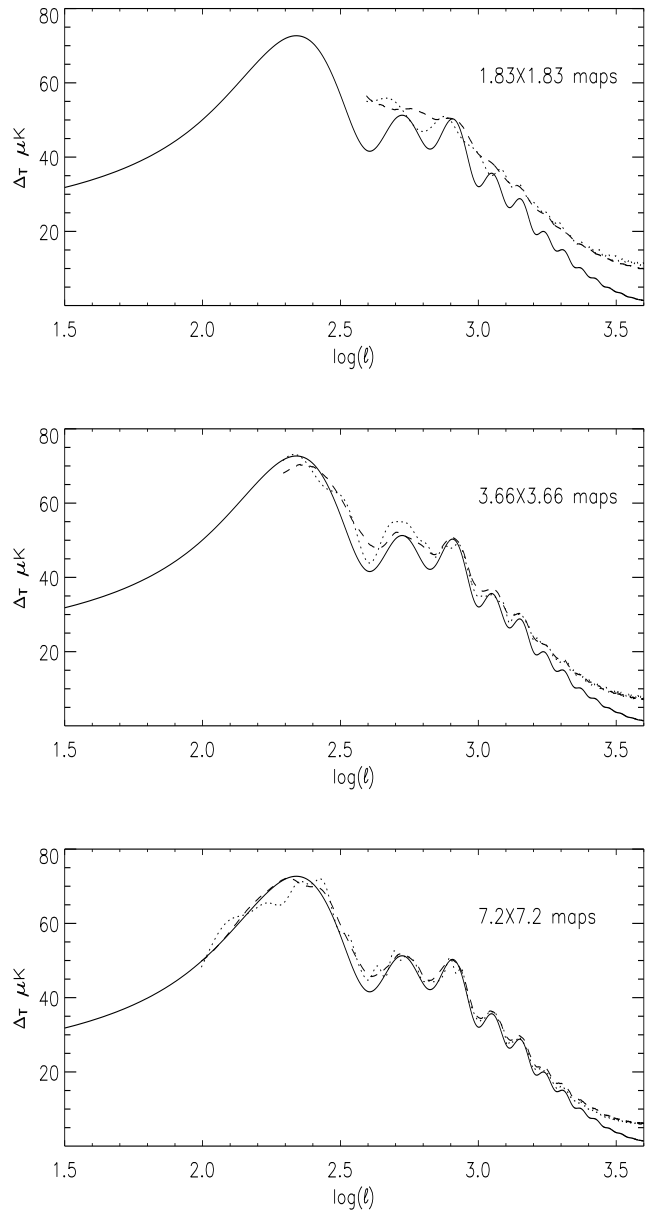,height=16cm,width=8.4cm}
%\vspace{0.5cm}
%% to leave a blank space
\caption{CMB power spectrum, $\Delta_{_{T}} $, in $\mu K$, as a function 
of $\log (\ell )$ in all panels. All the solid lines exhibit the CMB spectrum 
given by CMBFAST which is used as an input to simulate CMB maps.
Top, middle, and bottom panels correspond to 
$ 1.83^{\circ} \times 1.83^{\circ} $, $ 3.66^{\circ} \times 3.66^{\circ} $,
and $ 7.2^{\circ} \times 7.2^{\circ} $ maps, respectively. In the 
three panels, dotted and dashed lines are the spectra extracted
from the simulated maps by using two distinct estimators (see text).
}
\label{figu6}
\end{figure}

The spectra obtained from our small 
non-Gaussian RS maps have systematic errors
which are now approximately estimated. In order to 
perform this estimation, 
we begin with the analysis of Gaussian maps of the same size and,
then, results are considered as an approximated estimate 
of the errors associated to the non-Gaussian RS simulations. 
This procedure seems to be reasonable taking into account that
deviations from Gaussianity of the RS maps do not seem to be dramatic 
(see our discussion in section 5).
As it is well known, the relative errors of the $C_{\ell }$ quantities 
extracted from a Gaussian map is $\Delta C_{\ell } /C_{\ell } = [2/(2 \ell +1)]^{1/2}
f_{sky}^{-1/2} $, where $f_{sky} $ is the ratio between the map area 
and that of the full sky \citep{sco94, kno95}. 
According to this formula, the relative errors 
corresponding to $ 3.66^{\circ} \times 3.66^{\circ} $ Gaussian 
maps range between 
$ \sim 2.5 $ and $\sim 0.5 $ as $\ell $ takes on the values of the interval
$(500, 10000) $.  For
$ 1.83^{\circ} \times 1.83^{\circ} $ 
($ 7.2^{\circ} \times 7.2^{\circ} $ ) maps these errors increase
(decrease) by a factor $2$.  
These relative errors are so large that one could 
think that  
Gaussian maps of dominant anisotropy (and also  
our simulated maps) are not useful at all; however,
when $C_{\ell } $ quantities are extracted from one of the
mentioned Gaussian  
maps, the errors lead to a oscillatory angular spectrum
which changes from $\ell $ to $\ell +1 $; in other 
words, the resulting spectrum has high frequency oscillations 
which can be eliminated by means of suitable averages.
These averages are performed during (after) the $C_{\ell }$
estimations when the method proposed in \citet{arn02} (ANAFAST)
is used. The final averaged spectra are meaningful 
in a certain
interval of $\ell $ values. In order to define this interval, 
$ 1.83^{\circ} \times 1.83^{\circ} $ maps of the CMB dominant anisotropy
have been simulated by using: the angular power spectrum given by
CMBFAST (in the model assumed here) and a map making method based
on the fast Fourier transform \citet{sae96}. The resulting maps are 
analyzed with two different power spectrum estimators. In the
top panel of Fig. \ref{figu6}, the solid line is the CMBFAST spectrum
used as input in the simulated maps, the dashed line is the spectrum
extracted from the maps by using the power spectrum estimator
described in \citet{arn02} and, finally, the dotted line
has been obtained using ANAFAST (see above) to analyze an
equatorial healpix map obtained from our uniformly pixelised 
one. Both estimators give very similar spectra which can be 
considered as a rather good estimate of the true spectrum for
$500 \leq \ell \leq 2000$. Maps are too small to get the
spectrum for $\ell < 500$ and, furthermore, 
for $\ell > 2000$, the signal is too weak and, then, numerical and
statistical noise become comparable to it. 
The middle and bottom panels of Fig. \ref{figu6} have the
same structure as the top panel; however, in the middle (bottom)
panel the angular size of the maps is 
$ 3.66^{\circ} \times 3.66^{\circ} $ 
($ 7.2^{\circ} \times 7.2^{\circ} $). Comparing the panels 
of Fig. \ref{figu6}, one concludes that the more extended the maps, 
the better the resulting spectra  and the wider the $\ell $ interval 
where the spectra are good enough. 
In the case of the RS effect studied with HR simulations,
the size of our simulation boxes ($L=256 \ Mpc$)
and, consequently, the angular size of the resulting maps 
($ 1.83^{\circ} \times 1.83^{\circ} $) cannot be 
freely increased because any significant increment would lead to a 
problematic decrease of the 
N-body simulation resolution, nevertheless, 
if the above results corresponding to Gaussian maps are 
applied --as an approximation-- in the  
presence of deviations from Gaussianity, our 
$ 1.83^{\circ} \times 1.83^{\circ} $
maps should suffice to get a rough, but useful, estimate of the 
RS spectrum 
in a wide $\ell$ interval. This conclusion is confirmed by a 
direct study of RS maps presented in the next paragraph.

\begin{figure}
\psfig{file=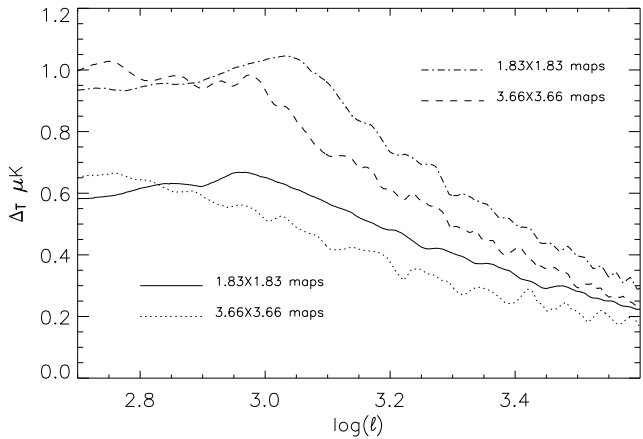,width=8.4cm}
%\vspace{0.5cm}
%% to leave a blank space
\caption{RS power spectrum, $\Delta_{_{T}} $, in $\mu K$, as a function 
of $\log (\ell )$. Dotted-dashed and dashed (solid and dotted) lines
have been obtained with LR (IR) simulations for $L=256 \ Mpc$
and $L=512 \ Mpc$, respectively. The angular sizes of the maps
are given inside the panel. Scales are smaller than $60 \ Mpc$ in all
cases.
}
\label{figu7}
\end{figure}

In order to have direct information about the quality of the 
spectra associated to $ 1.83^{\circ} \times 1.83^{\circ} $ RS maps,
two tests have been performed. In the first one, maps of this angular 
size have been obtained from LR simulations with $L=256 \ Mpc$
and, then, the angular power spectrum extracted from these maps (which is 
presented in the dashed-dotted line of Fig. \ref{figu7}) is compared with
that obtained from LR simulations with $L=512 \ Mpc$ which lead to 
$ 3.66^{\circ} \times 3.66^{\circ} $ RS maps (dashed line of the same
Figure). Since both simulations have the same resolution, differences
in the resulting spectra are due to the difference between the angular
size of the maps. The comparison of the dashed-dotted and dashed lines
of Fig. \ref{figu7} shows that, in the $ 1.83^{\circ} \times 1.83^{\circ} $ 
maps, the spectrum is slightly underestimated for $500 < \ell <800$ and 
overestimated
for $\ell >800$. The second test is identical to the first one except for
the resolution of the involved simulations. 
The $ 1.83^{\circ} \times 1.83^{\circ} $
maps are obtained with IR simulations and $L=256 \ Mpc$, 
and the maps whose angular size is
$ 3.66^{\circ} \times 3.66^{\circ} $ are created with IR simulations
in boxes with $L=512 \ Mpc$;
the corresponding spectra are displayed by the solid and dotted lines of 
Fig. \ref{figu7}, respectively. The comparison of these spectra indicates that,
for the smallest maps, the resulting spectrum is underestimated (overestimated)
for $500 < \ell <700$ ($\ell >700$). Furthermore the maxima of the 
dashed-dotted and solid lines appear as a result of the overestimations and
they disappear in the case of our most extended maps (see the dotted and 
dashed lines). Unfortunately, HR simulations cannot be used
to perform a third test similar to the previous ones (too great
computational cost of HR simulations with $L=512 \ Mpc$), nevertheless, 
previous discussion strongly 
suggests that, also in the case of $ 1.83^{\circ} \times 1.83^{\circ} $ RS
maps generated with HR simulations, the spectrum (dotted line 
of Fig. \ref{figu4})
should be underestimated and overestimated as in the first and
second tests (see the weak maximum observed in that dotted line); 
therefore, for $\ell > 800$, the $\Delta_{_{T}} $ quantities corresponding to 
$ 3.66^{\circ} \times 3.66^{\circ} $ HR maps should be smaller than those
extracted from $ 1.83^{\circ} \times 1.83^{\circ} $ HR maps
(dotted line of Fig. \ref{figu4}). 
In other words, the smallness of the maps leads to an overestimation 
of the angular power spectrum for large $\ell $ values.

Previous discussion about the 
$\Delta_{_{T}} $ uncertainties associated to 
both N-body resolution (Fig. \ref{figu4})
and map size (Fig. \ref{figu7}) can be summarized as
follows:
(1) for $\ell $ values between $700$ and $800$, 
the spectrum $\Delta_{_{T}} $ slightly depends on the
map size (see Fig. \ref{figu7}); however,
its dependence on N-body resolution (see Fig. \ref{figu4})
is important;
hence, the $\Delta_{_{T}} $ value corresponding to 
$\ell \simeq 700 $  
essentially depends on resolution.
From the visual analysis of Fig. \ref{figu4} one
easily concludes that
this value should be a number between $0.4 \ \mu K$ and
$0.5 \ \mu K$, 
(2) for $\ell > 800$, the overestimations of $\Delta_{_{T}} $
associated to N-body resolution and map size are comparable
and they add to generate a total excess of power which might
be close to $0.2 \ \mu K$. This number follows from a qualitative 
study of Figs. \ref{figu4} and \ref{figu7},
(3) on account of points (1) and (2), 
the dotted line of Fig. \ref{figu4} appears to be  
an upper limit to the RS spectrum, but it can be 
considered, in practice, as an useful rough estimation
of the true spectrum. The most important point is that the resulting 
spectrum has 
been obtained from fully nonlinear dark matter evolution 
(without approximating techniques o models), and  
(4) more resolution and more extended maps are necessary to
get a more accurate RS spectrum.

\section{Statistical Analysis}

Various reduced correlation functions are estimated and
then deviations from Gaussianity are analyzed. 
The correlations associated to n directions (or
n points on the last  scattering surface) are estimated for
$n \leq 6$. Given a map of the variable $\zeta $, and $m$ directions, 
the reduced angular correlations are:
\begin{equation}
C_{m} = \langle \zeta (\vec {n}_{1}) \zeta (\vec {n}_{2}) 
\cdots \zeta (\vec {n}_{m}) \rangle \ ,
\label{corr}
\end{equation}
where the average is over many statistical realizations.

\begin{figure}
\psfig{file=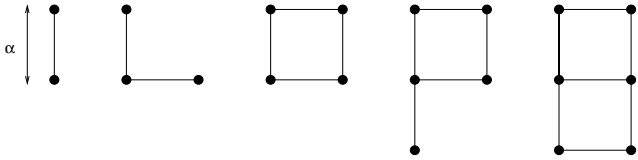,width=8.4cm}
%\vspace{0.5cm}
%% to leave a blank space
\caption{m-direction configurations for statistical analysis}
\label{figu8}
\end{figure}

Correlations depend on the relative positions of the 
chosen directions; hence, for CMB and RS maps, 
they depend on the figures that 
these directions draw on the last scattering surface. In this
paper we have used the sets of directions 
displayed in Fig.~\ref{figu8}, where 
the basic correlation angle, $\alpha$, is that
subtended by the two directions of the case $n=2$.
In Gaussian statistics it is well known that
\citep{pee80}: (i) all the 
$C_{m}$ correlations vanish for odd $m$ and, (ii)
for even $m$, all the correlations can be written in terms of 
$C_{2}$. Hereafter, these even Gaussian correlations 
are denoted $C_{gm}$ (the suffix g stands for Gaussian).
For the configurations of Fig.~\ref{figu8}, one easily 
obtains the relation:
\begin{equation}
C_{g4}(\alpha) = 2C^{2}_{2}(\alpha) + C^{2}_{2}(\sqrt{2} \alpha) \ ,
\label{cg4}
\end{equation}
leading to $C_{g4}(0) = 3C^{2}_{2}(0)$ for $\alpha = 0$, and the equation:                                              
\begin{eqnarray}
\nonumber
C_{g6}(\alpha) & = & 3C^{3}_{2}(\alpha) + 
2C_{2}(\alpha) C^{2}_{2}(\sqrt{2} \alpha) \\
\nonumber
& + & 4C_{2}(\alpha) C_{2}(2 \alpha) C_{2}(\sqrt{2} \alpha)\\
\nonumber 
& + & 2C_{2}(\sqrt{5} \alpha) C^{2}_{2}(\sqrt{2} \alpha)+
C_{2}(\alpha) C^{2}_{2}(\sqrt{5} \alpha) \\
& + & 2C_{2}(\sqrt{5} \alpha) C^{2}_{2}(\alpha)+
2C_{2}(\alpha) C^{2}_{2}(2 \alpha) \ , 
\label{cg6}
\end{eqnarray}
which takes on the form 
$C_{g6}(0) = 15C^{3}_{2}(0)$ for $\alpha = 0$.
Our RS simulated maps are not Gaussian and
we are interested in deviations from Gaussianity.
Even correlations can be used to estimate these
deviations as follows: (a) extract 
the $C_{2}$, $C_{4}$, and $C_{6}$ correlations from the 
simulated maps, (b) use the extracted $C_{2}$ correlation
to calculate $C_{g4}$, and $C_{g6}$ functions 
assuming Gaussianity, namely, using Eqs. 
(\ref{cg4})--(\ref{cg6}) and, (c) compare 
$C_{4}$ and $C_{6}$ correlations with functions 
$C_{g4}$ and $C_{g6}$, respectively. 
Differences between $C_{4}$ and $C_{g4}$ as well as
between $C_{6}$ and $C_{g6}$ measure deviations 
from Gaussianity. Furthermore,
odd correlations vanish in the Gaussian case
and, consequently, if at least one of the correlations $C_{3}$ 
and $C_{5}$ (for the configurations of Fig.~\ref{figu8})
appears to be different from zero,  we can conclude that 
the analyzed RS maps
are not Gaussian.

\begin{figure}
\psfig{file=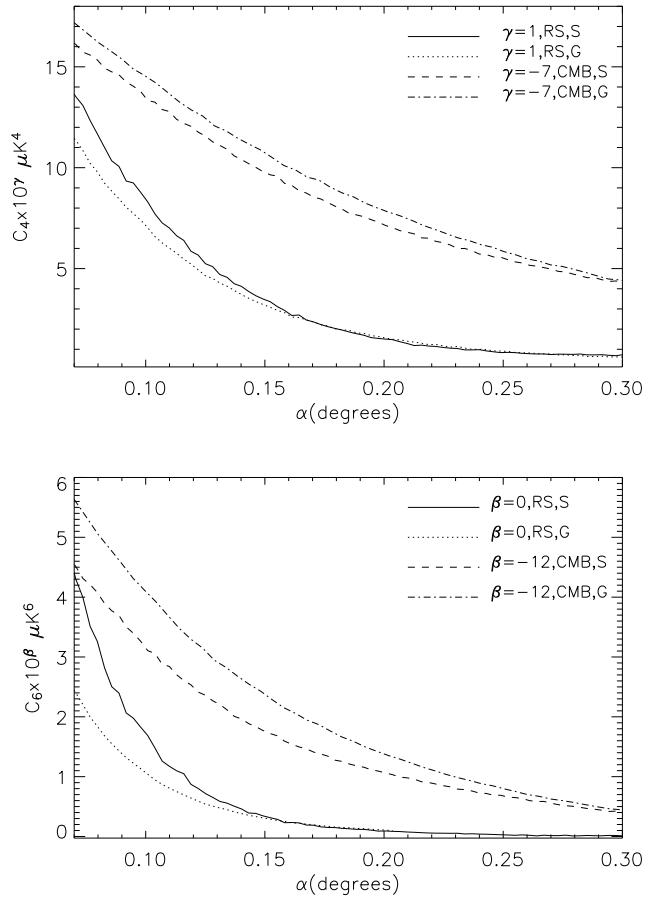,width=8.4cm}
%\vspace{0.5cm}
%% to leave a blank space
\caption{Even correlation functions obtained from
$ 1.83^{\circ} \times 1.83^{\circ} $ maps 
in terms of the correlation 
angle $\alpha $.  
$C_{4}$ ($C_{6}$) correlations are multiplied by an appropriate 
power of ten and given in $\mu K^{4} $ ($\mu K^{6} $).
Top (bottom) panel corresponds to
$C_{4}$ ( $C_{6}$ ) functions.
Dashed-dotted (dashed) line exhibits either $C_{g4}$  
or $C_{g6}$ ( $C_{4}$ or $C_{6}$ ) correlations 
extracted from CMB maps.
Solid (dotted) line shows either $C_{g4}$
or $C_{g6}$ ( $C_{4}$ or $C_{6}$ ) correlations 
found from RS maps. Each type of line is defined inside the
panel giving: (i) the power of ten exponent, (ii)
the type of map (CMB or RS), and (iii) the type of correlation
(S, for $C_{4}$ and $C_{6}$, and G, for $C_{g4}$ and $C_{g6}$)
}
\label{figu9}
\end{figure}

In order to get the $C_{m}$ correlation from a set of maps, 
the m-configuration of Fig.~\ref{figu8} is randomly placed 
on the maps and, then, the temperature contrasts in the nodes 
of the configuration are calculated by appropriate
interpolations in our maps (making use of their pixelisation); finally, 
the average of Eq. (\ref{corr}) is calculated considering
all the maps and all the locations of the m-configuration.
As it occurs in the case of the $C_{\ell} $ quantities,
the fact that our simulated maps are small implies 
that the $C_{m} $ correlations extracted from them are not the true ones.
In spite of this fact, it is now proved that 
the methods proposed 
above to detect deviations from Gaussianity --based on the use of even 
correlations-- work. In order to do that, these methods are first 
applied to CMB maps --Gaussian by construction-- having either the same
sizes as our simulated RS maps or larger ones; in this way, small  
Gaussian maps are characterized and, then, deviations of the RS maps 
with respect to this characterization point out a certain level of
non-Gaussianity.

\begin{figure}
\psfig{file=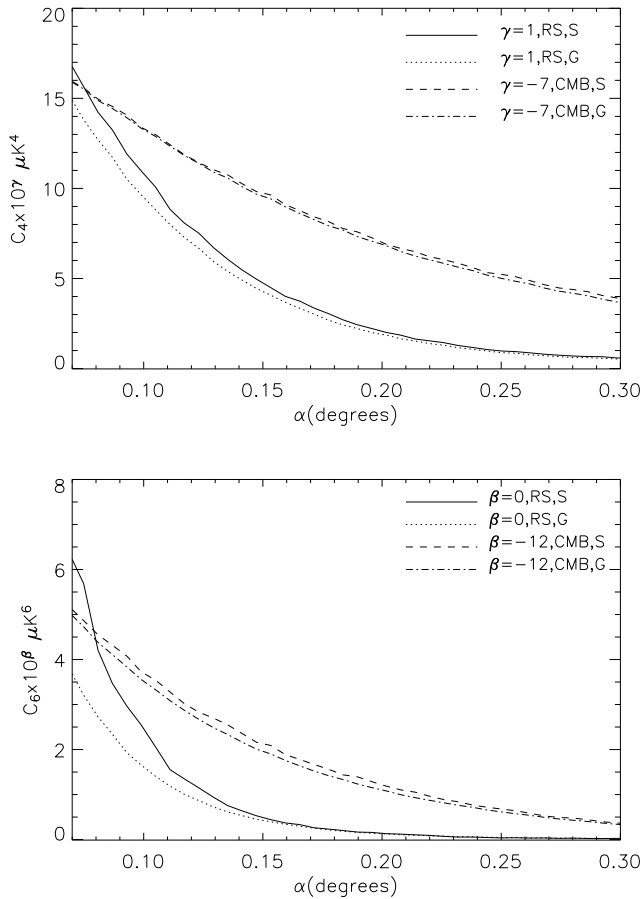,width=8.4cm}
%\vspace{0.5cm}
%% to leave a blank space
\caption{Same as in Fig.~\ref{figu9} for 
$ 3.66^{\circ} \times 3.66^{\circ} $ maps
}
\label{figu10}
\end{figure}

We begin with $ 1.83^{\circ} \times 1.83^{\circ} $ Gaussian CMB maps. 
Some correlations extracted from these maps are displayed
in the panels of Fig.~\ref{figu9}.
The dashed (dashed-dotted) line of the top panel exhibits the
$C_{4} $ ($C_{g4} $) correlations (see above), whereas the   
same lines of the bottom panel display
$C_{6} $ and $C_{g6} $ correlations.
Since the $ 1.83^{\circ} \times 1.83^{\circ} $ maps are too
small, correlations extracted from them are not the 
true correlations (which could be obtained from 
a large set of big maps) and, consequently, in spite of Gaussianity, 
functions $C_{4} $ and $C_{g4} $ slightly separate
(compare dashed and dashed-dotted lines in the top panel of
Fig.~\ref{figu9}). The same occurs for the correlations 
$C_{6} $ and $C_{g6} $  (compare dashed and dashed-dotted lines 
in the bottom panel of
Fig.~\ref{figu9}). 
In the case of $ 3.66^{\circ} \times 3.66^{\circ} $ maps,
the correlations are presented in 
Fig.~\ref{figu10} using the same notation as in
Fig.~\ref{figu9}. From Fig.~\ref{figu10} one easily 
concludes that
dashed and dashed dotted lines are much closer than in the 
corresponding panels of Fig.~\ref{figu9}. It has been verified that for
$ 7.2^{\circ} \times 7.2^{\circ} $ maps, the corresponding lines are
closer than in Fig.~\ref{figu10}. This behaviour characterize 
Gaussian maps, in which the $C_{4} $ and $C_{6} $ correlations 
approach the $C_{g4} $ and $C_{g6} $ functions, respectively, as 
the map size increases and the extracted correlations approach 
more and more the true ones. It is worthwhile to notice that,
although the correlations $C_{2} $, $C_{4} $ and
$C_{6} $ obtained from small maps are not the true correlations,
they allow us to verify the Gaussian behaviour 
described above, which can be applied to detect deviations 
from Gaussianity.

Even correlations extracted from 
$ 1.83^{\circ} \times 1.83^{\circ} $ and 
$ 3.66^{\circ} \times 3.66^{\circ} $ RS maps
are now analyzed. In order to do that, 
the study described in the last paragraph 
(for CMB maps) is
repeated for RS maps. Results are also presented in 
Figs.~\ref{figu9} and
\ref{figu10}; in which
solid and dotted lines are used to represent the
same correlations given by the dashed-dotted 
and dashed lines of the same Figures (see above). 
From the comparison of the 
solid and dotted lines one easily concludes that:
(i) for $ 1.83^{\circ} \times 1.83^{\circ} $ maps
(Fig.~\ref{figu9}), these lines 
separate for $0.07^{\circ} < \theta < 0.15^{\circ} $.
Since this separation is comparable to that 
corresponding to Gaussian CMB small maps (dashed-dotted and
dashed lines), it does not point out 
any deviation from Gaussianity, and (ii) 
for $ 3.66^{\circ} \times 3.66^{\circ} $ maps
(Fig.~\ref{figu10}), the separation between 
solid and dotted lines is much greater than that 
of the CMB Gaussian maps with the same size
(dashed dotted and dashed lines of Fig.~\ref{figu10}).
It occurs for all the angular scales we have 
represented, which include the scales 
corresponding to $500 < \ell < 2000$, for which
ou best estimation of the  RS spectrum (dotted line of 
Fig.~\ref{figu4}) has been obtained. This 
result (great separation) proves that the RS maps 
are not Gaussian because,
in the case of Gaussianity, the separation 
of the solid and dotted curves corresponding to 
$ 3.66^{\circ} \times 3.66^{\circ} $ maps
must be smaller than the separation 
corresponding to $ 1.83^{\circ} \times 1.83^{\circ} $ maps
(Gaussian behaviour defined above).

Odd correlations have been also considered. 
Maps of $ 1.83^{\circ} \times 1.83^{\circ} $ and
$ 3.66^{\circ} \times 3.66^{\circ} $ squared degrees 
have been used to study these correlations. 
In the case of CMB Gaussian simulations, the general
method (see above) for estimating
functions $C_{3} $ and $C_{5} $ has been applied using $n$   
small CMB maps. Different $n$ values
have been considered.
Results do not converge as $n$ increases because the true odd 
correlations vanish and, consequently, the functions $C_{3} $ and $C_{5} $
extracted from $n$ simulated Gaussian CMB maps are pure errors for any $n$.
This complete absence of convergence characterizes small Gaussian 
maps for any size. 
In the case of RS small maps with any of the above sizes, 
it has been verified that
the $C_{3} $ and $C_{5} $ correlations extracted from a number $m$ of
these maps do not converge as m-increases (as it occurs for Gaussian
maps); this means that we have not been able 
to detect the small odd correlations associated to the RS effect
(see next section for more discussion).

\section{Results and Discussion}

In this paper, an improved version of the numerical methods 
used in \citet{ali02} is applied to study the RS effect. Our main
improvements: preferred direction and appropriate cutoff were
described in \citet{cer04} and successfully applied in \citet{ant05}. 
These new elements make
negligible the magnification of the RS and lens effects 
due to the spatial periodicity of
the simulated universe, which is covered by identical simulation boxes.
The present size of these boxes must range between $\sim 256 \ Mpc$ and 
$\sim 512 \ Mpc$. Only in this way, good N-body simulations 
with large enough resolutions can be obtained 
using our PM code; as a result of this size restriction, 
we are constrained to work with small maps of the RS effect
having sizes between 
$1.83^{\circ} \times 1.83^{\circ} $ and
$3.66^{\circ} \times 3.66^{\circ} $ (see section 3).
Since the correlations obtained 
from these small maps are expected to have important uncertainties,
an exhaustive study about the accuracy and significance of these
correlations is developed. Furthermore,
methods specially designed for detecting deviations with 
respect to Gaussianity (using the correlations of 
small maps)
are described and applied.

The analysis of small CMB Gaussian maps of dominant anisotropy 
(CMBFAST angular power spectrum, see section 4) has been basic
in order to study both power spectrum estimations and 
non-Gaussianity detection. The following studies have been developed:
(1) the $C_{\ell }$ quantities extracted from Gaussian CMB maps
(for different sizes) have been compared 
with those used in their simulation; in this way, the  interval 
($\ell_{1}$,$\ell_{2}$)
where the recovered angular power spectrum is good enough (from a 
qualitative point of view) has been
estimated (see Fig.~\ref{figu6} of section 4), the existence of 
this interval has been interpreted 
taking into account the concept of sample variance and, 
(2) the method described in 
section 5 to look for deviations with respect to Gaussianity --from even 
correlations-- has been applied to analyze CMB maps with distinct 
sizes and, thus, Gaussian maps have been characterized, which has
been basic for 
further researches about deviations from Gaussianity. 
The interval ($\ell_{1}$,$\ell_{2}$) depends on the 
map size (see section 4) and the $C_{\ell} $ quantities have been
calculated for various of these sizes.  
The larger the size of the
RS map, the wider this interval. More extended maps are necessary to
get the angular power spectrum for $\ell < \ell_{1}$, and more
extended maps and greater resolutions are needed to find this spectrum 
in the case
$\ell > \ell_{2}$; namely, for very small angular scales.
There is the opinion that the RS effect could produce some 
feature in the angular power spectrum at small enough
angular scales, by this reason, 
our main project in this field is the use of P3M and AP3M N-body  
codes to get resolutions as high as possible in large enough boxes. 
In spite of this plan,  
we are skeptical about the existence of the mentioned feature
because the RS effect is produced by the time derivative of the
peculiar potential, which seems to be rather well estimated with 
the effective resolution of
$1 \ Mpc$ corresponding to the dotted line of Fig.~\ref{figu4}, in 
which, there is no any trace of incipient
features at large $\ell $ values; nevertheless, 
it must be recognized that,
for these large values, the uncertainties in our spectrum are 
maximum (superimposition of the overestimations
due to resolution and map smallness) and, consequently,
a moderated excess of power due to fully non-linear evolution 
could be hidden by our current uncertainties.  
Our method (based on N-body simulations) is fully nonlinear and,
moreover, it takes into account both the effect due to the gravitational 
collapse of cosmological structures and that produced by the motion 
of these structures in the box. These are the most promising aspects of 
the proposed techniques whose future applications 
--based on better N-body simulations and computers--  
can improve on both the
estimation of the RS spectrum (pointing out posible new features)
and the detection of deviations from Gaussianity. In other
words, although improvements on the applications of the ray tracing
procedure used here are possible, current
applications lead to too small 
RS maps and moderate N-body resolutions and, 
consequently, only an upper limit to the RS spectrum (rough estimate
of it) has been obtained in this paper. 
Fortunately, we are going to see that
this estimate suffices to discuss 
detection of the RS effect with WMAP and PLANCK satellites
(see below) and, furthermore,
our detection of deviations from Gaussianity in RS maps seems 
to be doubtless. 

A few comments about previous results are now necessary
for comparisons.
The angular power spectrum of the Rees-Sciama 
effect was studied by \citet{tul96} and \citet{sel96}.
In both papers, the effect was estimated for open and flat 
cold dark matter 
universes without cosmological constant. In the first of
these papers, the authors used N-body simulations and
a certain ray-tracing procedure to get the $C_{\ell} $
coefficients for $\ell < 700$. Among their results 
we emphasize the following ones:
($\alpha $) in the case of a flat universe (with h=0.75 and
$\sigma_{8} = 1$),
and for $\ell $ close to
$700 $, quantity $\Delta_{_{T}} $ takes on values around
$0.27 \ \mu K$, and ($\beta $) for $\ell \sim 100$ this quantity reaches a
maximum value close to $0.85 $. In the second paper, 
\citet{sel96} did not use any ray tracing procedure.
He used N-body simulations (second order perturbation theory)
to evolve the density contrast of strongly (mildly) non-linear structures and,
then, from the resulting contrast, the author estimated the power 
spectrum of the quantity 
$\partial \phi (\vec{x},t) / \partial t $. Finally, from this spectrum, the 
$C_{\ell} $ quantities were calculated.
Four cases corresponding
to different values of the parameters $\sigma_{8}$ 
and $\Omega_{m}h$ were studied using N-body simulations. 
Some results corresponding to these four cases are now 
summarized: (i) the temperature contrast
$(\Delta T /T)$ takes on values between
$10^{-6}$ and $10^{-7}$, (ii) quantity $\Delta_{_{T}} $
is maximum between $\ell = 100 $ and $\ell = 300 $,
(iii) the maximum values of $\Delta_{_{T}} $ 
range from $0.38 \ \mu K$ to $2.25 \ \mu K $ and, (iv) the values 
corresponding to $\ell \sim 700 $ ranges from 
$0.27 \ \mu K$ to $1.2 \ \mu K $ (the first of these values
corresponds to a flat universe with h=0.5 and $\sigma_{8} = 0.6$).
More recently, the RS effect has been estimated, in 
the concordance model, for various slightly different values
of the involved parameters (see, e.g. \citet{coor02} 
and \citet{hu02}); these estimations are not based on 
N-body simulations and they lead to small values of  
$\Delta_{_{T}} $ for $\ell \simeq 700$; these values are 
$\sim 0.17 \ \mu K$ in \citet{coor02} and 
$\sim 0.2 \ \mu K$ in \citet{hu02}.
As it has been discussed in section 4, the value of $C_{700} $
extracted from our small RS simulations seems to be between 
$0.4 \ \mu K$ and $0.5 \ \mu K $;
this approximated value seems to be a little greater than 
those found (with other models and methods) by
\citet{tul96}, \citet{sel96}, \citet{coor02}, and \citet{hu02}.
The $\Delta_{_{T}} $ spectrum given in the three last 
of these references and that obtained by us 
slowly decrease for $\ell > 700 $; nevertheless, our spectrum 
decreases slower than that shown in \citet{coor02} and \citet{hu02}.  
Our $\Delta_{_{T}} $ overestimations, at large 
$\ell $ values, associated to map smallness (see section 4)
seem to explain this difference.

The $C_{3} $ correlation induced by the RS effect was studied 
by \citet{mol95} and \citet{mun95}.
The main conclusion of these 
authors was that this odd correlation is 
smaller than its cosmic variance, which suggests that 
it is 
too small to be detected 
from a unique complete realization of the RS sky.
Of course, we have not any map of the full sky, but 
very small maps
of the RS effect and, consequently, it should not be 
possible the estimation of $C_{3} $ from 
these small parts of the full sky.
In spite of these considerations and taking into account that 
the mentioned 
estimates of $C_{3} $ are based on 
perturbation theory, which 
does not properly describe strong non-linear evolution 
inside cluster and substructures, 
we have tried the estimation of the $C_{3} $ and 
$C_{5} $ correlations from our
$1.83^{\circ} \times 1.83^{\circ} $ and
$3.66^{\circ} \times 3.66^{\circ} $ RS maps; nevertheless,
this estimate has not been possible, which confirms the low 
level of these odd correlations in the maps of the RS effect. 

\begin{figure}
\psfig{file=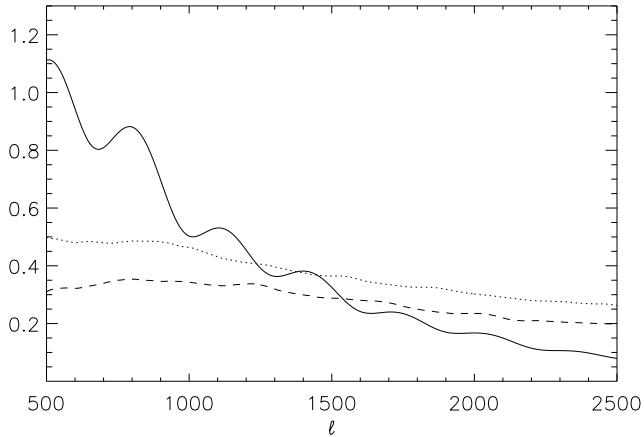,width=8.4cm}
%\vspace{0.5cm}
%% to leave a blank space
\caption{numbers in the vertical axis are temperatures in 
$\mu K$. The solid line displays the errors of
$\Delta_{_{T}} $ due to cosmic variance and the dotted
(dashed) line is the spectrum $\Delta_{_{T}} $ obtained from 
HR simulations 
for $\sigma_{8} =0.93$ ($\sigma_{8} =0.76$)
}
\label{figu11}
\end{figure}
Even correlations have been calculated for both CMB and RS 
small maps and, after comparisons, a certain deviation with respect 
to Gaussianity
has been pointed out in the RS case (see section 5); however, 
no deviations have
appeared in the analysis of the superimposition of RS 
and CMB maps. It is due to the fact that 
the RS non-Gaussian part is much smaller than the CMB Gaussian one.
The superimposition is almost Gaussian and, consequently,
cosmic variance errors, $\Delta C_{\ell} = [2/(2\ell + 1)]^{1/2} 
C_{\ell}$, are a good approximation to the $C_{\ell}$ 
uncertainty in full sky measurements.
These errors can be easily calculated taking into account that
the angular power spectrum of the superimposition is almost
identical to that of pure CMB. In the 
$\ell $ interval [500,2000], where we have found our best 
estimate of the RS spectrum (see section 4), one easily finds
the errors of $\Delta_{_{T}} $, in $\mu K$, associated to cosmic
variance. They are displayed in the solid line of Fig. 
\ref{figu11}. In the same Figure, the dotted line 
is our best estimate of the RS spectrum (also represented in the 
dotted line of Fig. \ref{figu4}), and the dashed line is a
new spectrum which has been obtained using the same method
as in the case of the dotted line, but assuming $\sigma_{8} = 0.76$
in our HR simulations. This new normalization is suggested 
by the analysis of the three year WMAP data 
published \citep{spe06}
after the submission of the first version of this paper.
In Fig. \ref{figu11} we see that,
for the two $\sigma_{8} $ values we have considered, our
upper limits to the RS spectrum are well below the cosmic errors
for $\ell < 1200 $; hence, taking into account that 
WMAP cannot accurately measure $C_{\ell} $ coefficients 
for $\ell > 1200 $ (see \citet{hin06}), one easily concludes that
the RS effect cannot be observed with this satellite.
Finally, the upper limits to the RS spectrum displayed in  
the dotted and dashed lines are well above $\Delta_{_{T}} $
errors for $1500 < \ell < 2000$; however, taking into account that 
the true RS spectra are 
expected to be around $0.2 \ \mu K$ smaller than 
our upper limits 
for large $\ell $ values (see section 4), it follows that 
the true $\Delta_{_{T}} $ values are not expected to be greater than 
the cosmic errors given by the solid line of Fig. \ref{figu11}.
This means that, although all the multipoles 
up to $\ell \simeq 2500$ could be measured with PLANCK, 
the RS effect would be too small to be detected with
this satellite. All these 
considerations strongly suggest that the RS effect will not be directly 
detected in the near future. Indirect detection based on correlations
of the CMB with the cluster distribution producing the effect 
deserves attention.

Simulations in great 
boxes would lead to more extended 
RS maps (improving on  power spectrum estimation and 
non-Gaussianity detection). 
In the case of 
large simulation boxes with $L \simeq 1000 \ Mpc $, which 
can be currently considered with appropriate N-body codes 
and computers, the magnification of the RS effect associated 
to periodicity is not 
negligible because eight of these boxes are yet 
necessary to cover the 
photon trajectory from $z=5.2$ to present time. Of course,
our method based on 
a preferred direction also applies for these big boxes. 
After minimum changes, our codes will be ready to built up better
RS maps from
appropriate N-body simulations, and also to statistically analyze these
maps.

\section*{Acknowledgments}
This work has been              
supported by the Spanish MEC
(project AYA2003-08739-C02-02 partially funded with FEDER) and
also by the Generalitat Valenciana (grupos03/170).  
Calculations were carried out at the 
Centro de Inform\'atica de la Universidad de Valencia 
(CERCA and CESAR).

\bsp

\label{lastpage}


\begin{thebibliography}{99}

\bibitem[\protect\citeauthoryear{Afshordi, Lho \& Strauss}{2004}]{afs04}
Afshordi N., Lho Y-S., Strauss M.A., 2004, Phys. Rev. 69D, 3524
\bibitem[\protect\citeauthoryear{Aliaga, Quilis \& S\'aez}{2002}]{ali02} 
Aliaga A.M., Quilis V., S\'aez D., 2002, MNRAS, 330, 625
\bibitem[\protect\citeauthoryear{Ant\'on et al.}{2005}]{ant05}
Ant\'on L., Cerd\'a-Dur\'an P., Quilis V., S\'aez D., 2005, ApJ, 628, 1
\bibitem[\protect\citeauthoryear{Arnau, Aliaga \& S\'aez}{2002}]{arn02}
Arnau J.V., Aliaga A.M., S\'aez D., 2002, A\&A, 382, 1138
\bibitem[\protect\citeauthoryear{Bennet et al.}{2003}]{ben03}
Bennet C.L. et al., 2003, ApJS, 148, 1
\bibitem[\protect\citeauthoryear{Boughn \& Crittenden}{2004}]{bou04}
Boughn S.P., Crittenden R.G., 2004, Nat, 427, 45
\bibitem[\protect\citeauthoryear{Cerd\'a-Dur\'an, Quilis \& S\'aez}{2004}]{cer04}
Cerd\'a-Dur\'an P., Quilis V., S\'aez D., 2004, Phys. Rev., 69D, 043002
\bibitem[\protect\citeauthoryear{Cooray}{2002}]{coo02} 
Cooray, A., 2002, Phys. Rev. 65D, 103510
\bibitem[\protect\citeauthoryear{Cooray}{2002}]{coor02} 
Cooray, A., 2002, Phys. Rev. 65D, 083518
\bibitem[\protect\citeauthoryear{Crittenden \& Turok}{1996}]{cri96} 
Crittenden, R.G., Turok, N., 1996, Phys. Rev. Lett., 76, 575 
\bibitem[\protect\citeauthoryear{Eke, Cole \& Frenk}{1996}]{eke96} Eke, V., 
Cole S., Frenk C.S., 1996, MNRAS, 282, 263
\bibitem[\protect\citeauthoryear{Fosalba, Gazta\~{n}aga \& Castander}{2003}]{fos03}
Fosalba P., Gazta\~{n}ga, E., Castander F., 2003, ApJ, 597, L89
\bibitem[\protect\citeauthoryear{Fosalba \& Gazta\~{n}aga}{2004}]{fos04}
Fosalba P., Gazta\~{n}ga, E., 2004, MNRAS, 350, 37
\bibitem[\protect\citeauthoryear{Fullana \& S\'aez}{2000}]{ful00} 
Fullana M.J., S\'aez D., 2000, New Astronomy, 5, 10
\bibitem[\protect\citeauthoryear{G\'orski, Hivon \& Wandelt}{1999}]{gor99}
G\'orski, K.M., Hivon, E., Wandelt, B.D., 1999, Proceedings of the 
MPA/ESO Conference on Evolution of Large Scale Structure, Eds.
Banday, A.J., Sheth, R.K., \& Da Costa, L., [astro-ph/9812350]
\bibitem[\protect\citeauthoryear{Hinsaw et al.}{2006}]{hin06} 
Hinsaw, G. et al., 2006, astro-ph/0603451   
\bibitem[\protect\citeauthoryear{Hockney \& Eastwood}{1988}]{hoc88}
Hockney R.W., Eastwood J.W., 1988,
Computer Simulations Using Particles (Bristol: IOP Publisings) 
\bibitem[\protect\citeauthoryear{Hu \& Dodelson}{2002}]{hu02} 
Hu W., Dodelson S., 2002, Annu. Rev. Astron. Astrophys., 40, 171
\bibitem[\protect\citeauthoryear{Jain, Seljak \& White}{2000}]{jai00}
Jain B., Seljak U., White S., 2000, ApJ, 530, 547
\bibitem[\protect\citeauthoryear{Knox}{1995}]{kno95}
Knox L., 1995, Phys. Rev., 52D, 4307
\bibitem[\protect\citeauthoryear{Mart\'{\i}nez-Gonz\'alez, Sanz \& Silk}{1990}]
{mar90}Mart\'{\i}nez-Gonz\'alez E., Sanz J.L., Silk J., 1990, ApJ, 355, L5
\bibitem[\protect\citeauthoryear{Mollerach et al.}{1995}]{mol95}
Mollerach S., Gangui A., Lucchin F., Matarrese S., 1995, ApJ, 453, 1
\bibitem[\protect\citeauthoryear{Munshi, Souradeep \& Starobinski}{1995}]{mun95}
Munshi D., Souradeep T., Starobinski A.A., 1995, ApJ, 454, 552
\bibitem[\protect\citeauthoryear{Nolta et al.}{2004}]{nol04}
Nolta et al., 2004, ApJ, 608, 10
\bibitem[\protect\citeauthoryear{Padmanabhan et al.}{2005}]{pad05}
Padmanabhan, N., Hirata, C.M., Seljak, U., Schlegel, D.J.,
Brinkmann, J., Schneider, D.P., 2005, Phys. Rev., 72D, 043525
\bibitem[\protect\citeauthoryear{Peebles}{1980}]{pee80} Peebles P.J.E.,
1980, The Large Scale Structure of the Universe, Princeton University press,
Princeton   
\bibitem[\protect\citeauthoryear{Quilis, Ib\'a\~{n}ez \& S\'aez}{1998}]{qui98}
Quilis V., Ib\'a\~{n}ez J.M., S\'aez D., 1998, ApJ, 502, 518
\bibitem[\protect\citeauthoryear{S\'aez, Holtmann \& Smoot}{1996}]{sae96}
S\'aez D., Holtmann E., Smoot G.F., 1996, ApJ, 473, 1
\bibitem[\protect\citeauthoryear{Scott, Sredniki \& White}{1994}]{sco94}
Scott D., Sredniki M., White M., 1994, ApJ, 421, L5
\bibitem[\protect\citeauthoryear{Seljak}{1996}]{sel96}
Seljak U., 1996, ApJ, 463, 1  
\bibitem[\protect\citeauthoryear{Seljak \& Zaldarriaga}{1996}]{selz96} 
Seljak U., Zaldarriaga M., 1996, ApJ, 469, 437  
\bibitem[\protect\citeauthoryear{Spergel et al.}{2006}]{spe06} 
Spergel, D.N. et al., 2006, astro-ph/0603449   
\bibitem[\protect\citeauthoryear{Tuluie, Laguna \& Anninos}{1996}]{tul96}
Tuluie R., Laguna P., Anninos P., 1996, ApJ, 463, 15
\bibitem[\protect\citeauthoryear
{Vielva, Mart\'{\i}nez-Gonz\'alez \& Tucci}{2006}]{vie06}
Vielva, P., Mart\'{\i}nez-Gonz\'alez, E., Tucci, M., 2006, MNRAS, 365, 891
\bibitem[\protect\citeauthoryear{White \& Hu}{2001}]{whi01}
White M., Hu W., 2001, ApJ, 537, 1
\end{thebibliography}
\end{document}